# Machine Learning Approaches for Detecting the Depression from Resting-State Electroencephalogram (EEG): A Review Study

Milena Čukić Radenković[1,2], PhD and Victoria Lopez Lopez[3*], PhD

[1]Department for General Physiology and Biophysics, University of Belgrade, Belgrade, Serbia
[2]Amsterdam Health and Technology Institute, HealthInc, Amsterdam, the Netherlands
[3]Facultad de Informatica, Universidad Complutense, Madrid, Spain

**Author Info**
[*] Correspondence concerning this article should be addressed to Victoria Lopez Lopez, Profesor J. Santesmases, 9, 28040 Madrid, Tel: +34629975771, Email: vlopezlo@ucm.es

Milena Čukić, Ph.D., Koningin Wilhelminaplein 644, 1062KS Amsterdam, the Netherlands, Tel: +31615178926, Email: milena.cukic@gmail.com or micu@3ega.nl

**Abstract**

In this paper, we aimed at reviewing present literature on employing nonlinear analysis in combination with machine learning methods, in depression detection or prediction task. We are focusing on affordable data-driven approach, applicable for everyday clinical practice, and in particular, those based on electroencephalographic (EEG) recordings. Among those studies utilizing EEG, we are discussing a group of applications used for detecting the depression based on the resting state EEG (detection studies) and interventional studies (using stimulus in their protocols or aiming to predict the outcome of therapy). We conclude with a discussion and review of guidelines to improve the reliability of developed models that could serve improvement of diagnostic and more accurate treatment of depression.

1. Introduction

Before 2030 depression can become the first reason for disability worldwide (WHO, 2014 and 2017; Mathew and Loncar, 2006). Only 11-30% of patients diagnosed with depression are reaching remission within the first year of treatment (Rush et al., 2008; Rost et al., 2002; Cipriani et al., 2009). Among other problems hampering the improvement is so-called co-morbidity, where a person can be diagnosed with more than one disorder, according to DSM-IV manual presently in use (or two persons can be labelled with the same disorder with completely non-overlapping symptoms). Unlike many other specializations in medicine, psychiatry is not using objective physiological tests in the diagnostics process (Shorter and Fink, 2010; Gillan et al., 2017). Many researchers and clinicians are aware that this diagnostic process needs improvement. Matching the patients with interventions, finding specific biomarkers and various technical solutions can provide much-needed improvement of this process.

Computational Psychiatry is a discipline conceived as a combination of computational neuroscience and psychiatry, with an aim to find neuro-biological underpinnings behind clusters of clinical symptoms, making it easier to adjust the treatment to the patients on a personal level (Montague et al., 2011; Wang and Krystal, 2014; Yahata et al., 2017). One of the most impressive studies in detecting three depression sub-types was performed by Tokuda and his colleagues, by analyzing demographic data, prior medical histories and MRI data (Tomoki Tokuda et al., 2018). There are two approaches to computational psychiatry: theory-driven and data-driven. A data-driven approach is typically using some kind of machine learning and seems to be much more reliable than the other approach due to comparably lover costs of data collection. Although the most popular work being published in last period applying the data-driven approach by utilizing MRI or fMRI data, the drawbacks of that approach are the subject of the debate among researchers. Much more realistic version of this approach, according to our opinion, would be relying on electroencephalographic (EEG) data, due to much lower costs and higher accessibility for patients. EEG is the oldest form of neuroimaging which is non-invasive and is very well based in neurology and neuroscience; in psychiatry, it is used to confirm the existence of epileptiform only. In comparison to fMRI for example, the time needed for recording and the price of processing, EEG is much more suitable for frequent testing. Data mining is, according to Witten and Frank (2005) 'the extraction of implicit, previously unknown, and potentially useful information from the data', and nowadays popular machine learning is one part of that discipline. Typical pipeline in this framework includes recording the EEG, managing the artefact removal (manual, software, or using artefact-free epochs), linear or nonlinear analysis of EEG, feature extraction, feature selection and then application of machine learning (both training and testing phase) method of choice. As highly structured data EEG (a matrix of voltage values as columns/recorded from different electrodes and time) is highly suitable for machine learning (Craik et al., 2019).

Physiological complexity is another area of research, still considered to be novel among medical professionals. It is based on Complex systems dynamic theory (popularly known as 'Chaos theory') and comprises of vast families of different approaches of analyses in a mathematical sense. Many researchers are utilizing those methods since physiological signals are known to be nonlinear, nonstationary and generated from a highly complex system which is usually operating far from the equilibrium state. To apply any mechanistic approach (suitable for stationary signals) for analysis of electrophysiological data - which are 3N: nonlinear, nonstationary and noisy (Klonowski, 2007) is yielding risks of wrong interpretation. Recent research showed that mathematical link exists between usually applied Fourier analysis and Fractal analysis (Kalauzi et al., 2012) and Fourier seems to be redundant for the former. Professor Klonowski speculates that the reason for omnipresent classical spectral analysis in electrophysiology is due to the deeply rooted utilization of it in medicine; nonlinear analysis is usually known just in research (Klonowski, 2007). For review about application of varying nonlinear methodologies in detecting the depression based on EEG see de la Torre-Luque (2017).

In the last ten years, the number of research utilizing some form of machine learning on EEG dataset to detect the depression or predict the outcome of treatment, related to depression is booming. This study aims at reviewing that literature to make a cross-section useful in realizing what best practices are. We opted to focus on the combination of physiological complexity (application of nonlinear measures of analysis of EEG) and computational psychiatry data-driven approach since we believe that this combination can lead to faster improvement of current clinical practice in treating depression.

In this paper, we are firstly describing our Method (section 2) and eligibility criteria, then we described Detection (section 3) and Interventional studies (section 4), and finally (in section 5) we discuss the drawbacks and possible solutions for noticed shortcomings, conclusions as well as suggestions for future work.

2. **Methods**

This systematic review of the literature aimed at finding and comparing published studies that employed a combination of nonlinear (and spectral) methods of analysis in combination with various machine learning methods in the depression detection task. Therefore, we made inclusion criteria list, since we knew that many studies were published in the last decade. Since we followed the literature for quite some time, we set the start in 2011, and finalize the search in May 2019. Given the fast pace development in this area of research due to faster computers, cloud utilization and internet improved performances, we think that this period for inclusion is sufficient. We systematically searched Web of Science and PubMed databases on May 24. 2019. with the following combination of keywords:

('Data mining' OR 'machine learning') OR ('Support Vector Machines' OR 'Neural Networks' OR 'Logistic regression' OR 'K Nearest Neighbors' OR 'Naïve Bayes' OR 'Linear regression' OR 'genetic algorithm' OR 'Decision three' OR 'Random Forest') AND ('EEG' OR 'Electroencephalography') AND ('Depression' OR 'MDD').

Also, databases that index both fields including Springer, Scopus and ScienceDirect were searched for relevant literature, as well as Cornell repository Arxiv.org.

After original search yielded 197 papers, the authors reviewed all the titles and abstracts in order to decide which of those are in line with our search criteria (eligibility testing): a study published between 2011 and 2019; detection of depression or forecasting the outcome of depression treatment; sample comprising of patients diagnosed with depression (MDD) and Healthy controls; EEG dataset (preferably resting-state EEG); utilization of fractal and nonlinear analysis as features for machine learning; utilization of machine learning for task of detection the depression. After primary selection (in which we read all the publications independently), our sample comprised of 32 publications which we agreed to downsize on 26 after internal discussion and comparative analysis. After reading the entire text of each publication we decided to include 14 detection studies and 12 interventional studies. To summarize, we included EEG studies only, those which are published in last 8 years, those which employed task classification performed by humans using EEG signals (excluded were power-analyses only, non-human feature selection, or those with no end classification studies) which performed machine learning task aimed at detecting depression. There were lots of studies describing mobile phone applications and online collection of data (psychiatry going online) utilizing machine learning, but we reviewed that in other research work already (Llamocca et al., 2018).

Before this systematic search, we also made a list of characteristics of a study that we are going to compare and discuss to present the best practices and results. First, we compared the size of the sample, and only one intervention study was large enough to analyze the sample bigger than 100 participants (just one comprised of female subjects only). Since we opted to include EEG studies only, we divided them on resting-state EEG (employed in diagnostic) and those which used a kind of stimulus during the recording; our idea from the point of view of nonlinear analysis is strongly for analyzing resting-state records, because in much prior research it was showed that they are most information-rich (Goldberger et al., 20011 and 2006). Berman showed that in depression, ruminative activities can be detected in task-free recordings only (Berman et al., 2011). Studies also varied in the number of electrodes used for recording, as well as standards used.

Next stage comparison would be for the method used for preprocessing of the data; there are some which used standard sub-bands (although, there is not yet published data or evidence that dividing EEG into sub-bands has any physiological meaning, Baçar et al., 2011) and those which analyzed the broadband signal. Those which used any of reductionistic approaches like Fourier's analysis, or wavelet transform or cosine transform, and those which analyzed the raw data. Those which removed artefact manually (but probably introduced other sources of artefacts that way), or removed them with some software (automatically), or decided to analyze the epochs from artefact-free sections of recorded signal (meaning no removal of artefacts). Another point to discuss was how much filtering and preprocessing at all was performed, or whether researchers focused on any part of the spectral content of the signal. Also, we compared what sampling frequency was applied to raw signals, which is important for further interpretation of results.

Next level was to determine what kind of analysis was performed on previously pre-processed data and what the chosen features were for further machine learning.

Studies differed also in the way how they chose to extract the features or to select the features.

We also noted whether the internal and external cross-validation was performed (and reported) and whether the study would be possible to replicate. And lastly, we compared what methods of machine learning were utilized in every paper, as well what was the accuracy after the testing phase, what was the sensitivity and specificity. Another question was whether they used ROC curves to probe the goodness of performed accuracy. We tried to make exhaustive analysis to those publications which complied with our eligibility criteria.

3. **Diagnostic studies**

In the present literature, there are several approaches in examining the changes in the complexity of EEG characteristic for depression. Among researchers there seem to be appearing a consensus that the characteristic of depression is elevated the complexity of EEG when compared to healthy peers (for review see, de la Torre-Luque, 2017). From previously published fMRI and DTI (FA) studies (de Kwaasterniet et al., 2013; Vederine et al., 2011) and graph theory applications on EEG signal (Kim et al., 2013) we know that in different depressive disorders changes in functional connectivity are confirmed. That is possibly reflected on the excitability of cortex so we could detect the difference in EEG between people diagnosed with depression and healthy controls (Ahmadlou et al., 2011; Bachmann et al., 2013; Hosseinifard et al., 2014; Faust et al., 2014). The conclusion of de la Torre-Luque reviewing study (2017) was that 'EEG dynamics for depressive patients appear more random than the dynamics of healthy non-depressed individuals.' Also, there is a consensus about the utilization of more than one nonlinear measure, because different measures are detecting unique features of the EEG signals 'revealing information which other measures were unable to detect' (Burns and Rayan, 2015).

We are reviewing here 14 studies (which we classified as Detection studies) published between 2011 and 2019. The problem with this cohort of studies is similar like with those trying to elucidate the changes of complexity from depressed patient's EEG; the direct comparison is challenging due to very different methodologies used. One of the first studies using resting-state EEG to classify depressed persons and healthy controls were the one by Ahmadlou (Ahmadlou et al., 2012). They also aimed at comparing two different algorithms for calculating the fractal dimension. What they found is in line with the work of several other authors (Esteller et al., 1999 and 2001; Castiglioni, 2010). Ahmadlou and colleagues wanted to examine which of two algorithms for calculating the fractal dimension (FD) is better as a feature of resting-state EEG for classification of MDD from healthy adults: Katz FD or Higuchi FD (HFD). From previous literature, they knew that different algorithms for calculating FD may differently interpret self-similarity and irregularity of time series since they are calculating it in different ways. Esteller et al., (2001) showed that KFD is more robust to noise compared to Petrosian's and Higuchi's algorithm, but other showed (Raghavendra and Narayana 2009; Castiglioni, 2010) that KFD is dependent on the sampling frequency, amplitude and waveforms, which is a disadvantage in analyzing biosignals. The disadvantage of HFD is, according to Esteller (2001) that it is more sensitive to noise than KFD. The sample for this study comprised of 12 non-medicated MDD patients and 12 healthy controls. The DSM-IV and Beck depression scale was used for scores for MDD. In their experiment resting-state, EEG was recorded for 3 minutes, with closed eyes, and the sampling rate was 256Hz. They opted to record just frontal electrodes for this experiment (7 electrodes Fp1, Fp2, Fz, F3, F4, F7 and F8, 10/20 standard) because of the previous finding that prefrontal cortices showed abnormal functioning in MDD. They used wavelets to decompose the raw EEG signal into 5 standard sub-bands (gamma, beta, alpha, theta, and delta), but they also analyzed the broadband signal. Klonowski showed (2007) that wavelet is a little bit better than classical Fourier's decomposition, but still distorting signal under study. Ahmadlou and team used averaged calculated KFD and HFD values (they divided it between the left and right electrodes and averaged it) as features for Enhanced Probabilistic Neural Networks (EPNN) after application of ANOVA which evaluated the ability of a feature to discriminate the groups based o variations both between and within groups. Authors found that MDD and non-MDD are more separable in the beta band based on HFD (contrary to the previous belief that the differentiation is best in the alpha band) and that HFD in both beta and gamma bands in MDD is higher than in healthy participants. That implied on higher complexity of signal recorded from frontal cortices (according to their data left frontal lobe is more affected). Based on HFD (which performed better than KFD) they obtained high accuracy of 91.3%.

Subha Puthankattil and her colleagues (2012) aimed at a classification of EEG records of depressed patients and controls by utilization of relative wavelet energy (RWE) and artificial feedforward neural network. Their sample comprised of 30 depression patients and the age and gender-matched healthy controls (16 females, 14 males). There is no information on whether the patients were medicated or not. The recording was taken from four locations in total; FP1-T3 (left) and FP2-T4 (right hemisphere). For 5 minutes they recorded resting-state EEG (the information about closed or open eyes is missing), sampling frequency was 256Hz, and they used notch-filter, but also utilized total variation filtering (TVF) for high-frequency noise. The signal was divided into sub-bands, eight levels multiresolution decomposition method of discrete wavelet transform (DWT) was used. As for feature extraction, an RWE analysis provided information about the signal energy distribution at different decomposition levels; twelve features were extracted for training and testing NNs. Nine features included values of RWE for different frequency bands, and two were obtained by observing the trend of the variation of the average RWE of EEG signals. The signal energy RWE is higher in depression. Coif let 5 showed the highest correlation coefficient and indicated the best match for depression patient's EEG (out of an array of 23). The performance of artificial neural networks yielded an accuracy of 98.11% (normal and depression signals). Sensitivity was 98.7%; Selectivity was 97.5%; specificity was 97.5% (Puthankattil et al., 2012).

Hosseinifard and colleagues (2014) examined the nonlinear analysis of EEG in 45 unmedicated depressed patients (23 females; 20-50 years old) and 45 healthy controls (19 to 60 years old). DSM-IV interview and Beck Depression Inventory were used. Their study aimed to classify healthy and depressed persons and to improve the accuracy of classification. The resting-state EEG with eyes closed was recorded for 5minutes from 19 electrodes from 10/20 standard system. Sampling frequency was 256Hz, and they used high and low plus notch filtering. Artefacts were removed manually. They divided raw EEG signal into standard sub-bands and applied both classical (Welch method) and four different nonlinear measures (detrended fluctuation analysis (DFA) Higuchi fractal dimension (HFD), correlation dimension and Lyapunov exponent) to characterize the signal. After feature extraction was performed, one classical and four nonlinear measures were calculated for all 19 electrodes for each person. Each feature vector (19 features related to 19 electrodes) is applied to K-nearest neighbours (KNN), Linear Discriminant Analysis (LDA) and Linear Regression (LR) classifiers. Two third of the sample was used for the training phase and the remainder for the test set. Leave-one-out cross-validation (LOOCV) method was applied to the training dataset. For feature selection they used a genetic algorithm (GA) the size of the population is set to 50, cross over the rate to 80% (They also tried PCA, but GA outperformed it significantly). Classification accuracy was the best in the alpha band for LDA and LR both reaching 73.3% (the worst was KNN in delta and beta and LDA in the delta with 66.6%). The best accuracy in the experiment was obtained by LR and LDA classifiers. The accuracy of all classifiers increased when the signal was characterized with nonlinear features, not classical power (LR reached 90% with correlation dimension). The conclusion was that 'nonlinear features give much better results in the classification of depressed patients and normal subjects' contrary to the classical one. Also, they concluded that depression patients and controls differ in the alpha band more than other bands, especially in the left hemisphere (Hosseinifard et al., 2014).

Faust and colleagues used EEG signals in the automated detection of depression (Faust et al., 2014). They probably used the same sample as Subha D. Puthankattil, and Paul K. Joseph, namely 30 patients diagnosed with clinical depression and 30 controls (age from 20 to 50; 16 females and 14 males) recorded earlier (2011/12) in Calcut, Kerala, India. Only 4 electrode positions were used, left FP1-T3 and right FP2-T4. They also used wavelet packet decomposition (WPD, Db8 wavelet) to extract appropriate sub-bands from the raw signal. Those extracted sub-bands were input for calculating several entropy measures; Bispectral entropy (Ph, including Higher-order spectra HOS technique, from Fourier analysis), Renyi entropy (REN), Approximate entropy (AppEn) and Sample entropy (SampEn). The process of extraction the sub-bands comprised of sending the original data through a sequence of down-sampling and low pass filters which defined the transfer function (which is like classical spectra analysis distorting the information content of the data, according to Klonowski, 2007). Also, prior to that extraction, researchers claim that high-frequency components did not contribute relevant information (contrary to our findings, Čukić et al., 2018) and removed them as well. After applying Student's t-test to evaluate features, several classification algorithms were applied. Gaussian Mixture Model (GMM), Decision Trees (DT), K-nearest neighbors (KNN), Naïve Bayes Classifier (NBC), Probabilistic Neural Networks

(PNN), Fuzzy Sugeno Classifier (FSC) and Support Vector Machines (SVM) were used. In this experiment, the authors applied ten-fold stratified cross-validation. The accuracy was 99.5%, sensitivity 99.2%, and specificity 99.7%. Contrary to Hosseinifard they claim that the EEG signals from the right part of the brain discriminate better the depressive persons (Faust et al., 2014).

The sample used for another study (Acharya et al., 2015) comprised of 15 healthy and 15 depressed persons (20-50-year-olds). Resting state EEG was recorded from just four positions in 10/20 standard system, left FP1-T3 and right FP2-T4, during 5 minutes (eyes closed and eyes opened). The sampling rate was 256Hz, notch-filtered (also comprising from 2000 samples like in Faust and Puthankattil, the method and sample were strikingly similar), artifacts were manually removed. Feature extracting consisted of 15 different measures: fractal dimension (Higuchi fractal dimension, HFD), largest Lyapunov exponent (LLE), sample entropy (SampEn), DFA, Hurst's exponent (H), higher order spectra features (weighted Centre of bispectrum, W_Bx, W_By), bispectrum phase entropy (EntPh), normalized bispectral entropy (Ent1) and normalized bispectra squared entropies (Ent2, Ent3), and recurrence quantification analysis parameters (determinism (DET), entropy (ENTR), laminarity (LAM) and recurrent times (T2)). These extracted features are ranked using the t value. The information whether some of them were calculated on standard EEG sub-bands and other on broadband signal is missing (like classical spectral measure high order spectra utilizing Fourier's analysis, must have been computed in sub-bands, but that was not mentioned). After a large number of trials, the authors decided based on a comparison of values to formulate Depression Diagnosis Index taking into account only LAM, W_By and SampEn, without the explanation. It says that 'DDI is a unique formula that yields nonoverlapping ranges for normal and depression classes.' This (probably) heuristically obtained index is used here instead of usually utilized classifiers. Features are ranked based on t value and fed to classifiers one by one obtaining the accuracy higher than 98%, sensitivity higher than 97% and specificity more than 98.5%. This best result is reportedly obtained by utilization of SVM with a polynomial kernel of order 3 (for both left and right hemisphere; they used averaged values for left and right hemisphere), although in previous papers by the same authors SVM as discarded before. The text has an ambiguity in 'features are fed to SVM classifier' and in the next sentence 'SVM classifier yielded the highest classification performance with the average accuracy…'

(Acharya et al., 2015). Whether SVM is actually used is among other inconsistencies in this groups' work report. In comparison to Faust and colleagues used EEG signals in the automated detection of depression (Faust et al., 2014). They probably used the same sample as Subha D. Puthankattil, and Paul K. Joseph, namely 30 patients diagnosed with clinical depression and 30 controls (age from 20 to 50; 16 females and 14 males) recorded earlier (2011/12) in Calcut, Kerala, India. Only 4 electrode positions were used, left FP1-T3 and right FP2-T4. They also used wavelet packet decomposition (WPD, Db8 wavelet) to extract appropriate sub-bands from the raw signal. Those extracted sub-bands were input for calculating several entropy measures; Bispectral entropy (Ph, including Higher-order spectra HOS technique, from Fourier analysis), Renyi entropy (REN), Approximate entropy (AppEn) and Sample entropy (SampEn). The process of extraction the sub-bands comprised of sending the original data through a sequence of down-sampling and low pass filters which defined the transfer function (which is like classical spectra analysis distorting the information content of the data, according to Klonowski, 2007). Also, prior to that extraction, researchers claim that high-frequency components did not contribute relevant information (contrary to our findings, Čukić et al., 2018) and removed them as well. After applying Student's t-test to evaluate features, several classification algorithms were applied. Gaussian Mixture Model (GMM), Decision Trees (DT), K-nearest neighbors (KNN), Naïve Bayes Classifier (NBC), Probabilistic Neural Networks (PNN), Fuzzy Sugeno Classifier (FSC) and Support Vector Machines (SVM) were used. In this experiment, authors applied ten-fold stratified cross-validation. The accuracy was 99.5%, sensitivity 99.2%, and specificity 99.7%. Contrary to Hosseinifard they claim that the EEG signals from the right part of the brain discriminate better the depressive persons (Faust et al., 2014).

The sample used for another study (Acharya et al., 2015) comprised of 15 healthy and 15 depressed persons (20-50-year-olds). Resting state EEG was recorded from just four positions in 10/20 standard system, left FP1-T3 and right FP2-T4, during 5 minutes (eyes closed and eyes opened). The sampling rate was 256Hz, notch-filtered (also comprising from 2000 samples like in Faust and Puthankattil, the method and sample were strikingly similar), artifacts were manually removed. Feature extracting consisted of 15 different measures: fractal dimension (Higuchi fractal dimension, HFD), largest Lyapunov exponent (LLE), sample entropy (SampEn), DFA, Hurst's exponent (H), higher order spectra features (weighted Centre of bispectrum, W_Bx, W_By), bispectrum phase entropy (EntPh), normalized bispectral entropy (Ent1) and normalized bispectra squared entropies (Ent2, Ent3), and recurrence quantification analysis parameters (determinism (DET), entropy (ENTR), laminarity (LAM) and recurrent times (T2)). These extracted features are ranked using the t value. The information whether some of them were calculated on standard EEG sub-bands and other on broadband signal is missing (like classical spectral measure high order spectra utilizing Fourier's analysis, must have been computed in sub-bands, but that was not mentioned). After a large number of trials, the authors decided based on a comparison of values to formulate Depression Diagnosis Index taking into account only LAM, W_By and SampEn, without the explanation. It says that 'DDI is a unique formula that yields nonoverlapping ranges for normal and depression classes.' This (probably) heuristically obtained index is used here instead of usually utilized classifiers. Features are ranked based on t value and fed to classifiers one by one obtaining the accuracy higher than 98%, sensitivity higher than 97% and specificity more than 98.5%. This best result is reportedly obtained by utilization of SVM with a polynomial kernel of order 3 (for both left and right hemisphere; they used averaged values for left and right hemisphere), although in previous papers by the same authors SVM as discarded before. The text has an ambiguity in 'features are fed to SVM classifier' and in the next sentence 'SVM classifier yielded the highest classification performance with the average accuracy…' (Acharya et al., 2015). Whether SVM is actually used is among other inconsistencies in this groups' work report. In comparison to Faust, who analyzed (probably) the same sample, here SampEn among other features is showing significantly higher values for depression than controls.

Bairy et al. (2015) used a discrete cosine transform (DCT) to decompose the raw EEG (of depressed and healthy persons) data to frequency sub-bands. Further, they calculated sample entropy, correlation dimension, fractal dimension, Lyapunov exponent, Hurst exponent and detrended fluctuation analysis (DFA) on DCT coefficients and the characteristics features are ranked by utilization of t-value. These features are used as input for classifiers DT, SVM, KNN, and NB. SVM with radial basis function (RBF) yielded an accuracy of 93.8%, the sensitivity of 92% and specificity of 95.9% (Bairy et al., 2015). We cannot say whether internal or external validation was performed, nor the details of, for example, the method used to calculate fractal dimension (that description is missing, so we cannot compare for example our work on Higuchi fractal dimension), hence the reproducibility of this study which claims so high accuracy is close to zero.

Another study is published in the same year (Mohammadi et al., 2015). This study used a sample of 53 MDD patients and 43 HC. EEG recordings lasted for 3 minutes (vigilance controlled eyes closed and eyes opened) in resting condition. The sampling rate was 500 Hz, a bandpass filter was used. Brain Vision Analyzer Software served for automatic artifact rejection. 28 electrodes are used for FFT analysis yielding a total of 12 datasets. After preprocessing the data, cleaning, and normalization, they applied Linear Discriminant Analysis (LDA) to map features into a new feature space (data evaluation phase). Authors applied Genetic Algorithm (GA) to identify the most significant features. They build predictive models with Decision Tree (DT). In two experiments, they first analyzed each frequency band individually, while the second experiment analyzed the bands together. The model showed an average accuracy of 80% (MDD vs. HC) (Mohammadi et al., 2015). There is no clear information about the checking of liability of their high accuracy nor both internal and external validation (in terms of a good generalization).

Liao et al. (2017) proposed a method based on scalp EEG and robust spectral-spatial EEG feature extraction based on kernel eigen-filter-bank common spatial pattern (KEFB-CSP). They first filter the multi-channel EEG signals (30 electrodes traces) of each sub-band from the original sensor space to new space where the new signals (i.e., CSPs) are optimal for the classification between MDD and healthy controls, and finally applies the kernel principal component analysis (kernel PCA) to transform the vector containing the CSPs from all frequency sub-bands to a lower-dimensional feature vector called KEFB-CSP (with 80% accuracy). Their sample comprises of 12 (4 males, 8 females, 43 -70 years old) patients plus 12 healthy controls, and resting state EEG was measured for 5 minutes. KNN, LDA, and SVM were used. They also used the majority voting strategy based feature selection.

Mumtaz and colleagues published their study on the same year as Liao: 2017; and another one in the same task in 2018. It relied on resting state EEG recorded from 33 MDD patients and 30 Healthy controls. Measures were spectral power of different frequency bands for diagnosing depression. A matrix was formed from extracted features, z-score standardization was applied, according to its mean and variance. To determine the most significant features a weight was assigned to each feature based on its ability to separate the target classes according to the ROC criterion. Only the most significant features were used for training and testing the classifier models: Logistic regression (LR), SVM and Naïve Bayesian (NB). The models are validated with the application of 10-fold cross-validation that has provided the metric for accuracy sensitivity and specificity. LR (acc 97.6%, sens 96.66%, spec 98.5%), NB (acc 96.8%, sens 96.6%, spec 97.02%), SVM (accuracy 98.4%, sensitivity 96.6% and specificity 100%). In their 2018 study the same group of authors used EEG-derived synchronization likelihood (SL) features for further machine learning: support vector machine (SVM), logistic regression (LR) and Naïve Bayesian (NB). This research resulted in SVM classification accuracy = 98%, LR classification accuracy = 91.7%, and NB classification accuracy = 93.6% .

Yet another study utilizing resting state EEG was published by Bachman et al., 2018. In this study (after the first one from 2013 when they explore the possibility of detection with novel spectral index SASI and Higuchi FD, but back then with k=50), Bachmann and colleagues performed classification task with resting state EEG measured from just one electrode. Their sample comprises of 13 depression patients (medication-free) and 13 healthy controls; they analyzed alfa power variability and relative gamma power, but also Higuchi's fractal dimension and Lempel-Ziv complexity. From 30 channel EEG, they opted on relying on a record from just one electrode. Features were used for classification by utilization of logistic regression with leave-one-out cross-validation. They reached maximal accuracy of 85% with HFD and DFA, but also HFD and LZC, and for only one nonlinear measure maximal 77%. This time, for calculating HFD they used k=8 (like in all of our former studies). Researchers claim that accurate detection of depression is possible with utilization the record from only one electrode and explain that with the possible acceptance from clinicians who prefer straightforward methods of detection. They also concluded that 'there is no single superior measure for detection of depression.'

In our study from 2018 (Čukić et al. 2018), we aimed to elucidate the effectiveness of two non-linear measures, Higuchi's Fractal Dimension (HFD) and Sample Entropy (SampEn), in detecting depressive disorders when applied on EEG. We recorded EEG in 21 participants diagnosed with depressive disorder and 20 healthy age-matched peers. The 10/20 International system for electrode placement was used (19 channels). The HFD and SampEn of EEG signals were used as features for seven machine learning algorithms including Multilayer Perceptron, Logistic Regression, Support Vector Machines with the linear and polynomial kernel, Decision Tree, Random Forest, and Naïve Bayes classifier, discriminating EEG between healthy control subjects and patients diagnosed with depression. We confirmed earlier observations that both non-linear measures can discriminate EEG signals of patients from healthy control subjects. Besides, our results suggest that if there is a proper feature selection, a useful classification is possible even with a small number of principal components. Average accuracy among classifiers ranged from 90.24% to 97.56%. Among the two measures, SampEn had better performance.

We concluded that using HFD and SampEn and a variety of machine learning methods we can accurately discriminate patients diagnosed with depression vs. controls which can serve as a sensitive, clinically relevant marker of depression. In comparison with previously mentioned studies which also used resting state EEG, we confirmed that the number of electrodes is important because from PCA readings it is clear that every electrode has its own contribution to the result (Čukić et al., 2018). And, an elevated complexity can be detected on all the positions (as also reported by Bachmann et al., 2013). Further, we cannot say that all mentioned studies are giving sufficient information for replication, like for example Bairy, who did not even state what method he used for calculating the fractal dimension, not to mention the algorithm. Others concentrated on the improvement of classification but did not undertake all the measures necessary to reach unwarranted optimism in their results. Last but not least all the studies (ours included, although we stated it was a pilot study) had very modest samples, which is affecting the generalizability of the model.

**Table 1:** Fourteen Detection Studies filtered with our inclusion criteria, published from 2011-2019. We compared sample sizes, basic technical details (the number of electrodes of recording rsEEG, sampling frequency), preprocessing details, features used for machine learning, machine learning models used and finally obtained accuracy in every study.

| Study | Sample (MDD + HC) | Electrodes, fs (Hz) | Preprocessing | Features | ML models | Accuracy |
|---|---|---|---|---|---|---|
| Ahmadlou et al 2012. | 12 +12 | 7, 256Hz | Wavelets and spectral bands (Fourier), bootstrap | Higuchi's and Katz's FD | Enhanced Probabilistic neural networks | 91.30% |
| Puthankattil et al, 2012 | 30(16m+14f) + 30 | 4, 256Hz | Wavelet , Total variation filtering, multiresolution decomposition | Wavelet entropy | RWE, artificial Feed Forward networks | 98.11% |
| Hosseinifard et al, 2014 | 45 + 45 | 19, 1KHz | Standard spectral bands | Power, DFA, Higuci, correlation dimension, Lapynov exp | KNN, LR, linear discriminant | 90% |
| Faust et al, 2014 | 30 + 30 | 4 (2 left, 2 right), 256Hz | Wavelet package decomposition | ApEn, SampEn, REN, bispectral phase entropy | PNN, SVM, DT, KNN, Naïve Bayes, GMM, Fuzzy Gueno Classifier | 99.50% |
| Bairy et al, 2015 | 30+30 (left brain only) | ? | Discrete cosine transform | SampEn, FD, CD, Hurst exp, LLE, DFA | DT, KNN, NB, SVM | 93.80% |
| Acharya et al., 2015. | 15+15 | 2 left, 2 right, 256Hz | broadband | FD, LLE, SampEn, DFA, H, W-Bx, W_By, EntPh, Ent1,DET, ENTR, LAM, T2 (DDI) | SVM, KNN, Naïve Bayes, PNN, DT | 98% |
| Mohammadi et al., 2015. | 53+43 | 28 (10/10), 500Hz | Standard bands/FFT, LDA, Genetic Algorithm (GA) | Spectral only | Decision tree (DT) | 80% |
| Puthankattil et al, 2014 | | | | | | |
| Liao et al., 2017 | 12+12 | 30, 500Hz | Common spatial pattern (CSP) | Spectral (CSP) | KEFB-CSB | 80% |
| Mumtaz et al., 2018 | | | | | | 87.50% |
| Mumtaz et al, 2017 | 33+30 | 19 (EO, EC), 256Hz | Fourier | Alpha interhemispheric asymmetry | LR, SVM, NB, | 98.40% |
| Mumtaz et al., 2018 | | | 10 fold cross validation | | | |
| Bachman et al, 2018 | 13+ 13 | 1, 1KHz | Fourier | HFD, DFA, Lempel-Ziv cmplx and SASI | Logistic regression | 88% |
| Čukić et al, 2018. | 26+20 | 19, 1KHz | Broadband EEG, 10 fold cross validation ,PCA | HFD+SampEN | MP, LR, SVM (with linear and polynomial kernel), DT, RF, NB | 97.50% |

## 4. Interventional EEG studies

There are also studies published in the same time interval (2011-2018) based on EEG registration, but the difference from work mentioned above, is that they opted to use a stimulus (hence, not resting state EEG), a sound stimulation, or ERP, so we will briefly mention their results here. Kalatzis and colleagues published the first study (our exception from time frame for chosen studies for review) about the SVM-based classification system for discriminating depression by using P600 component of ERP signals (Kalatzis et al., 2004). EEG was recorded on 15 electrodes, and a sample comprised of 25 patients and an equal number of healthy controls. The outcomes of SVM classification were selected by Majority vote engine (MVE). Classification accuracy reportedly was 94% when using all leads, and from 92% to 80% when using only right or left points for classification. They concluded that their findings support the hypothesis that depression is associated with the dysfunction of right hemisphere mechanisms mediating the processing of information that assigns a specific response to a particular stimulus. Lee et al., (2011) tried to predict the treatment response of major depressive disorder. Their study was designed to check whether the connectivity strength of resting state EEG could be a potential biomarker (ROC was 0.6 to 0.8) to answer this question. They concluded that '…the stronger the connectivity strengths, the poorer the treatment response.' The experiment also suggested that frontotemporal connectivity strengths could be a potential biomarker to differentiate responders and slow responders or non-responders in MDD. We tried to compare our results, but their sampling frequency is low as 100Hz, so that was difficult. Also, in a 2011 study,

Cavanagh and colleagues analyzed EEG recordings from 21 medication-free MDD patients and 24 healthy controls while performing probabilistic reinforcement learning task (Cavanagh et al., 2011). They measured EEG responses to error feedback, which can demonstrate selective alteration of avoidance learning, important in MDD. Khodayari-Rostamabad and colleagues probed machine learning methodology as a prediction model for a successful outcome of SSRI medication in MDD, based on resting-state EEG recorded prior the treatment (Khodayari-Rostamabad et al., 2010 and 2013). The sample comprised of 22 participants (11male, 11 female). For the experiment, they used only 16 electrodes (10/20 standard) in opened and closed eyes recording for 6.5 minutes and combined sections of it in six files per person. Welch model analysis yielded various spectral measures but mentioned 'only as candidate features' since they did not want to state in advance what feature would have the predictive power. After selecting the features extracted from EEG, authors fed them to the mixture of factor analysis (MFA) model, whose output is the predicted response in the form of a likelihood value; leave-one-out randomized permutation cross-validation procedure was used for validation. For visualization (but also a reduction of dimensionality) they used kernelized principal component analysis (KPCA). Authors did not perform evaluation on unseen sample, nor they compared the features with healthy controls, relying solely on spectral measures of their modest sample. They reported overall prediction accuracy of 87.9%.

A study from 2014, tried to predict the response of treatment in depression (Arns et al., 2014). The authors claimed that there is no difference between MDD and HC in non-linear EEG measures (they used Lempel-Ziv complexity), but somehow came to the conclusion that nonlinear measures are adding the value to this research; yet, they claim it is the first study to utilize nonlinear metrics to predict outcome of depression treatment (rTMS in their case). According to their reported method, the potential reason for that could be concentration on just one specific band and not on analysis of broadband signal, for many researchers after (and before) them succeeded to find significant difference by utilizing many nonlinear measures for this kind of detection task (Ahmadlou et al, 2012; Bachmann et al, 2013, 2018; Hosseinifard et al, 2013; Mumtaz et al, 2015; Mohammadi et al, 2015; Čukić et al., 2018). They also claimed that they were 'the first' to use complexity measures in this task. Nandrino and Pezard performed that approach to analysis of EEG in depression in 1994, as well as several other researchers groups (Nandrino and Pezard, 1994). Bachmann and colleagues (2018) applied exactly the same methodology (Lempel-Ziv complexity) and demonstrated significant differentiation between patients and controls. Mumtaz used in several papers spectral measures, but found the useful difference in predicting the outcome of treatment in depression (Mumtaz et al., 2017 and 2018).

Etkin and colleagues applied machine learning in the task of predicting the medication therapy outcome in MDD by utilizing cognitive testing (Etkin et al., 2015). They used pattern classification with cross-validation to determine individual patient-level composite predictive biomarkers of antidepressant outcome based on test performance and obtained 91% accuracy.

Erguzel and colleagues (Erguzel et al., 2016) tested their optimized classification methods on 147 participants with MDD treated with rTMS. They tested the performance of a genetic algorithm (GA) and a back-propagation neural network (BPNN); they were evaluated using 6-channel pre rTMS EEG patterns of theta and delta frequency bands. By using the reduced feature set, they obtained an increase under the receiver operating curve (AUC) of 0.904. Zhang et al. explored neural complexity in patients with post-stroke depression (Zhang et al., 2015) in a resting state EEG study. Their sample comprised of 21 post-stroke patients and 22 ischemic-stroke non-depression and 15 healthy controls: 16 electrodes were used for recording of resting state EEG. Lempel-Ziv complexity (LZC) was used to assess changes in complexity from EEG. PSD (depressed) patients showed lower neural complexity compared with PSND (non-depressed) and Control subjects in the whole brain regions. LZC parameters used for Post-stroke depression (PSD) recognition possessed more than 85% in specificity, sensitivity and accuracy suggesting the feasibility of LZC to serve as a screening indicator for PSD. Additionally, there were two antidepressive treatment non-response prediction studies: Shahaf et al. (2017), and al-Kaysi et al. (2017). Shahaf and his colleagues developed new electrophysiological attention-associated marker from a single channel (two electrodes: Fpz and one earlobe) using 1 min samples with auditory oddball stimuli and showed to be capable of detecting a treatment-resistant depression (26 patients, 10 controls). Al-Kayasi and team aimed at predicting tDCS treatment outcome of patients with MDD using automated EEG classification. They accurately predicted 8 out of 10 participants when using FC4-F8 (with accuracy 76%), and 10 out of 10 when using CPz-CP2 (accuracy 92%). This finding demonstrates the feasibility of using machine learning to identify patients to respond to tDCS treatment. Cai et al., 2018. Utilized only three electrodes on prefrontal positions to record the signal, while stimulating their participants with a sound. They claim that due to a small number of electrodes which can be easily positioned, their method has a great potential of translation to clinics. Cai and colleagues used electrophysiological database comprising of 92 depressed patients and 121 healthy controls; resting state EEG was recorded while sound stimulation (they used pervasive prefrontal lobe electrodes on positions Fp1, Fp2, and Fpz). After denoising (Finite Impulse Response, FIR filter) they combined Kalman derivative formula and Discrete Wavelet Transformation, and Adaptive predictor Filter; a total of 270 linear and nonlinear features were extracted (it is not clear what were they). Feature selection was minimal-redundancy-maximal-relevance, which reduced the dimensionality of the feature space. Four classification methods were applied: SVM, KNN, Classification threes and Artificial Neural Networks (ANN). For evaluation, they used 10-fold cross-validation. KNN had the highest accuracy of 79.27%. Jaworska and her colleagues published two papers; in 2018 and in 2019. First, they examined a variation of pre-treatment EEG in order to predict the treatment success in depression (Jaworska et al., 2018) and in second one they performed a machine learning study to predict the outcome of pharmacology treatments in 51 MDD patients in 12 week study (Jaworska et al., 2019). They used both electrophysiological and demographic data (including MADRS scores before and after the treatment) as well as source-localized current density (sLORETA) and utilized Random Forest for classification, with 78-88% accuracy depending of complexity of the model. They also used kernel principal component analysis to reduce and map important features. As many research mentioned above, this lays the groundwork for research on personalized, "biomarker"-based treatment approaches. For this group of studies, it can also be said that they are challenging to compare methodologically, but they are the part of the same effort of showing that not only detection but monitoring and predicting the pace of recovery, or output of the treatment (sometimes called 'responders' detection) is possible. The problem with both detection and interventional studies is generally modest samples and almost total absence of external validation process (on previously un-seen data, from an independent sample), which is putting in question typically high accuracies they reported.

## 5. On overrated optimism in machine learning and how the present methods can be improved to serve in clinical practice

To predict clinical outcomes or relapses (for example, after remission in recurrent depression) would be of great clinical significance especially in clinical psychiatry. However, developing a model for predicting a particular clinical outcome for the previously unseen individual have certain challenges both methodological and statistical (Whelan, 2013). A group of authors elucidated risks, pitfalls and recommend the techniques how to improve model reliability and validity in future research (Whelan and Garavan 2013; Gillan and Whelan, 2017; Yahata et al., 2017). The authors described that neuroimaging researchers who start to develop such predictive models are typically unaware of some considerations inevitable to accurately assess model performance and avoid inflated predictions (called 'optimism') (Whelan and Garavan, 2013; Cho et al., 2019). The common characteristics to that kind of research are: classification accuracy is typically overall 80-90%; the size of sample is typically small to modest (less than 50-100 participants); the samples are usually gathered on a single site. Support vector machines (SVM) and its variants are popular but recommendable is the use of embedded regularization frameworks, at least with absolute shrinkage and selection operator (LASO) (Yahata et al., 2017). Leave-one-out and k-fold cross validation are also popular procedures for validation (for model evaluation), and generalization capability of a model is typically untested on an independent sample (Yahata et al., 2017). For model evaluation or even reduction, Vapnik-Chevronenkis dimension should be used (Vapnik, 1988). A common denominator to majority of studies is a lack of external validation. In that sense, there is one study which did not use imaging, but demonstrated an impeccable methodology in machine learning in every aspect, in the task of prediction the responders to medication in MDD (Chekroud et al., 2016).

From a methodological point of view, there are many problems to resolve. What is the problem with a generalization? When we test the generalizability, we are basically testing whether or not a classification is effective in an independent (not shown to the algorithm before) population. When developing the model, one doesn't want to train the classifier on a general characteristic of a sample; for example, if using nonlinear measures, they can differ because some of the measures change with age (Goldberger et al., 2000) or they can be characteristic for gender (Ahmadi et al., 2013). Some authors call this 'nuisance variables', because the algorithm basically learns to recognize that particular dataset with all its characteristics. It is infamous overfitting and consequently the treatment of nuisance variables. Overfitting happens when 'a developed model perfectly describes the entire aspects of the training data (including all underlying relationships and associated noise), resulting in fitting error to asymptotically become zero' (Yahata et al., 2017). For that reason, the model will be unable to predict what we want on unseen data (test data). The size of the sample is usually small to modest (typically less than 100, in Chekroud is, for example, more than 4000, using collaborative dataset). Hence, the balancing of the complexity of the model against the sample size is essential for improving prediction accuracy for unseen (test) data (Yahata et al., 2017). How that goal can be achieved? By collecting more data. Like in collecting other more expensive neuroimaging data, it would be a solution to establish standard set-up and start collaborative projects, because one single site is usually not capable of gathering large samples alone. The model for this in EEG collections could be one of large collaborative projects like RDoC, STAR*D, IMAGEN, etc. Also, co-recording with fMRI and MEG should be a solution. Another option would also be to use wireless EEG caps. Another line of research is developing wireless EEG caps (Epoch, ENOBIO Neuroelectrics, iMotions, just to mention the few) which can be used for research in the environment without restraining the patient, and even for monitoring of recovering from severe episodes. If wireless EEG recorder would become accessible soon, we are sure that early detection and timely intervention will prevail fast. In the frameworks like National Institute of Mental Health Research Domain Criteria and European Roadmap for Mental Health Research which aim at finding stratifications that are based on biological markers that cut across current calcifications (Marquand et al., 2016) that should be possible. Then with big collaborative efforts, we may approach the conditions to extract genuinely reliable models for clearly defined neuromarkers for future clinical use (Gillan and Dow, 2017). Large-scale imaging campaigns and collection of general population data are the conditions for translation of those research findings to clinics. By allowing their usual check-up medical data to be a part of such organized collaborative efforts patients would also contribute to the improvement of this precise diagnostic of a (near) future. According to Wessel Kraiij (Data of value, 2017) P4 concept for healthcare improvement stands for: Prediction, Prevention, Personalization and Participation. An important motivation is the observation that healthcare is too focused on disease treatment and not enough on prevention. And another important observation (Kraiij, 2017) is, that treatment and diagnosis are based on population averages. In some cases, the treatment has a negative effect. So, there is a lot of room for precision improvement (and for other three Ps aside Personalization). The collection, analysis and sharing of the data play an important role in improvement of the healthcare that we know. In that sense, the first project to implement 4P is SWELL project part of a Dutch national ICT program COMMIT (between 2011 and 2016 in the Netherlands, Leiden University).

Whelan and Garavan (2013) precisely addressed other methodological issues, overfitting included. Their goal was to describe how regression models can appear, incorrectly, to be predictive, and to describe methods for quantifying and improving model reliability and validity. The authors conclude that '…perhaps counterintuitively to those who deal primarily with a general linear model, optimism increases as a function of the decreasing number of participants and the increasing number of predictor variables in the model (*model appears better as sample size decreases*)' (Whelan and Garavan, 2013).

Despite this expert knowledge about false optimism from our machine learning research, rare is that kind of attempt to quantifying the model performance. To resolve this, one could collect more data. The theory of data mining is clear; all those models work best on more significant numbers. Use the repository to test your developed model on the unseen cohort. At least, what I learned is that we need statistics to stop making fools from ourselves. Data mining is the art of finding the meaning from supposedly meaningless data (Peter Flack, 2014). A minimum rate of ten cases per predictor is common (Peduzzi et al., 1996), although not a universal recommendation (Vittinghof and McCulloch, 2007). Optimism can also be lowered with the introduction of the regularization term (Moons et al., 2004). Also, using previous information to constrain model complexity relying on Bayesian approaches is recommendable. Bootstrapping is another helpful method (Efron and Tibshirani, 1993) as well as cross-validation (Efron and Tibshirani, 1997). Cross-validation tests the ability of the model to generalize and involves separating the data into subsets. Both Kohavi and Ng described the technique (Kohavi et al., 1996; Ng et al., 1997). Besides, very useful and efficient ten-fold cross-validation, Elastic Net is useful to optimize parameters. Ng stated that '…optimism becomes unreliable as the probability of overfitting to the test data increases with multiple comparisons' (Ng et al., 1997). One can use several functions available in MATLAB (The MathWorks, Natick, Massachusetts) like lassoglm, bootstrap for bootstrap sampling, or several functions for Bayesian analysis, or crossvalind for testing sets and cross-validation. To conclude Whelan wrote on the importance of keeping training and test subsets completely separate; 'any cross-contamination will result in optimism' (Whelan and Garavan, 2013). Further research is needed to reframe nosology in psychiatry and improve patient's journey to remission.

**Acknowledgements**: Part of this work was supported by the project Social Big Data – CM ( S2015/HUM-3427)

**Conflict of interest statement**: The authors report no conflict of interests.